%% file: main.tex
\documentclass[sigconf]{acmart}
\usepackage{multirow}
\usepackage{rotating}
\usepackage{array}
\usepackage{float}
\usepackage{afterpage}
\usepackage{balance}

\copyrightyear{2026}
\acmYear{2026}
\setcopyright{cc}
\setcctype{by}
\acmConference[SIGCSE TS 2026] {Proceedings of the 57th ACM Technical Symposium on Computer Science Education V.1}{February 18--21, 2026}{St. Louis, MO, USA.}
\acmBooktitle{Proceedings of the 57th ACM Technical Symposium on Computer Science Education V.1 (SIGCSE TS 2026), February 18--21, 2026, St. Louis, MO, USA}
\acmISBN{979-8-4007-2256-1/2026/02}
\acmDOI{10.1145/3770762.3772660}

\begin{document}

\title[Competing or Collaborating? The Role of Hackathon Formats in Shaping Team Dynamics and Project Choices]{Competing or Collaborating? The Role of Hackathon Formats in Shaping Team Dynamics and Project Choices}


\author{Sadia Nasrin Tisha}
\orcid{0000-0003-4360-830X}
 \affiliation{%
   \institution{University of Maryland, Baltimore County}
   \country{}
   \city{Baltimore, MD, USA}}
   \email{stisha1@umbc.edu}

\author{Md Nazmus Sakib}
\orcid{0009-0003-8282-3931}
\affiliation{%
  \institution{University of Maryland, Baltimore County}
  \country{}
  \city{Baltimore, MD, USA}}
  \email{msakib1@umbc.edu}

 \author{Sanorita Dey}
 \orcid{0000-0003-3346-5886}
 \affiliation{%
   \institution{University of Maryland, Baltimore County}
   \country{}
   \city{Baltimore, MD, USA}}
 \email{sanorita@umbc.edu}
 








\renewcommand{\shortauthors}{Sadia Nasrin Tisha, Md Nazmus Sakib, Sanorita Dey}

\begin{abstract}
Hackathons have emerged as dynamic platforms for fostering innovation, collaboration, and skill development in the technology sector. Structural differences across hackathon formats raise important questions about how event design can shape student learning experiences and engagement. This study examines two distinct hackathon formats: a gender-specific hackathon (GS) and a regular institutional hackathon (RI). Using a mixed-methods approach, we analyze variations in team dynamics, project themes, role assignments, and environmental settings. Our findings indicate that GS hackathon foster a collaborative and supportive atmosphere, emphasizing personal growth and community learning, with projects often centered on health and well-being. In contrast, RI hackathon tend to promote a competitive, outcome-driven environment, with projects frequently addressing entertainment and environmental sustainability. Based on these insights, we propose a hybrid hackathon model that combines the strengths of both formats to balance competition with inclusivity. This work contributes to the design of more engaging, equitable, and pedagogically effective hackathon experiences.

\end{abstract}

\ccsdesc[500]{Human-centered computing~Empirical studies in collaborative and social computing}

\ccsdesc[300]{Human-centered computing~HCI theory, concepts and models}
\ccsdesc[500]{Human-centered computing~Empirical studies in HCI}

\keywords{Team Dynamics, Hackathon Projects, Collaboration, Distribution of Work}


\maketitle

\input{introduction_sd}

\input{related_works}

\input{Method_sd}

\input{result_sd}

\input{Discussion}

\bibliographystyle{ACM-Reference-Format}
\balance
\bibliography{sample-base}

\appendix

\end{document}

%% file: introduction_sd.tex
\section{Introduction}
Learning in computing education is a multifaceted process that extends beyond the traditional classroom boundaries. While formal education plays a foundational role, meaningful learning also occurs through informal and experiential platforms such as community projects, online initiatives, industrial training, and co-curricular activities~\cite{sefton2012learning}. Among these, hackathons have emerged as dynamic, time-bound events that facilitate hands-on computing education through collaboration, creativity, and rapid problem-solving~\cite{oyetade2024evaluation, araaejo2025embracing}.

Hackathons are short-term, intensive events where participants develop innovative solutions to real-world challenges~\cite{sotaquira2025hackathons, towhidi2022hackathons, servin2024hack}. Although initially rooted in software and programming communities, hackathons now span diverse domains, including gaming, healthcare, sustainability, education, and social good. These events provide a unique learning environment, enabling participants to apply technical and creative skills, collaborate in teams, and develop essential soft skills such as communication, leadership, and project management. A hallmark of hackathons is their ability to foster autonomy, allowing individuals to choose their projects, define objectives, and experiment with methodologies, promoting a deep sense of engagement and personal investment in computing education~\cite{brambilla2022hackathons}.
However, while hackathons are universally designed to encourage innovation and teamwork, the structure and composition of these events can significantly shape participants' experiences. Team formation rules, thematic focus, and demographic targeting often vary across events. Some hackathons are open to all, while others are specifically designed for targeted audiences, such as women, non-binary individuals, or novice coders, to promote equity and inclusion. These structural differences raise important questions about how learning, engagement, collaboration, and project outcomes vary across formats.

We, therefore, examine two distinct types of hackathon models organized by higher education institutions: gender-specific hackathons (GS), which focus on empowering underrepresented students (particularly women), and regular institutional hackathons (RI), which reflect a more traditional format without focusing on any specific student population. Our goal is to understand how the differences in formats and organizational structures influence participant engagement, collaboration, role distribution, and learning experiences. Specifically, we investigate how participants learn through peer interaction, mentorship, and active engagement and how these dynamics vary based on the hackathons' designs. Additionally, we explore how project ideas and themes differ between GS and RI hackathons, analyzing what these differences reveal about the broader goals, values, and outcomes encouraged by each format. Two key research questions guide this study:

\begin{enumerate}
\item \textbf{RQ1:} How do the environmental settings of gender-specific versus regular institutional hackathons influence students' learning experiences through collaboration, engagement, and perceived competitiveness?

 \item \textbf{RQ2:} Whether and how the format of a hackathon (gender-specific vs. regular-institutional) influences the project types presented by participants?
\end{enumerate}
 
Through a mixed-methods approach that includes semi-structured interviews, on-site observations, and analysis of the project archive, this study provides a comparative understanding of how different hackathon formats shape learning experience, participation, and engagement. Our analysis focuses on team dynamics, role distribution, project choices, and thematic orientations across different hackathon types. By integrating qualitative insights with quantitative techniques, the study reveals the nuanced ways in which hackathon design shapes computing education. Our findings offer actionable insights for designing more effective hackathon environments in computing and STEM education.

%% file: related_works.tex
\vspace{-0.08 in}
\section{Related work}
Hackathons have become widely recognized as effective platforms for informal learning, providing students and early-career technologists with practical experiences that enable them to apply their technical skills, collaborate in teams, and tackle real-world challenges. Unlike traditional classrooms, hackathons promote peer interaction, independence, and problem-solving, creating rich opportunities for learning and personal growth~\cite{nandi2016hackathons, paganini2021promoting}. Previous studies have highlighted hackathons as not only hubs of innovation and creativity but also as immersive environments for experiential education and the development of soft skills such as communication, leadership, and problem-solving~\cite{briscoe2014digital, araujo2025embracing, decker2015understanding}. These settings support both technical advancement and social development through observation, mentorship, and collaborative knowledge exchange. However, researchers have noted that the benefits of hackathons are not distributed equally among participants. Women and non-binary participants often face subtle but persistent barriers, including role stereotyping, reduced confidence, and limited access to mentorship or leadership opportunities~\cite{paganini2020female, richard2015stitchfest, kube2024hacking, paganini2021promoting}. These disparities are further reinforced in competitive or high-pressure settings where dominant norms may marginalize underrepresented voices~\cite{schauer2025internal, vitores2016trouble, frachtenberg2022underrepresentation}.

To counteract these imbalances, inclusive hackathon models have emerged, such as women-only or gender-diverse events that aim to cultivate safer, identity-affirming spaces through intentional design practices like gender-aligned mentorship, community-building activities, and proactive facilitation~\cite{kos2019understanding}. While these approaches have shown promise, most existing research focuses either on specific event case studies or on generalized barriers to participation, without systematically comparing how different hackathon structures impact participant experiences.
Our study fills this gap by directly comparing gender-specific and regular institutional hackathons to understand how event format influences participant engagement, peer interaction, role distribution, and project outcomes. By integrating qualitative and quantitative methods across two large-scale hackathons, we provide a nuanced understanding of how learning and collaboration unfold differently in gender-specific versus regular institutional environments. This comparative lens advances research and guides organizers to create inclusive and effective hackathons in computing and STEM education.

%% file: Method_sd.tex
\vspace{-0.08 in}
\section{Method}
In this study, we used a mixed method research design to thoroughly investigate the dynamics in two contrasting hackathon models. We chose a qualitative approach to capture detailed insights through direct observation and participant interviews. This section aims to provide a deeper understanding of the nuances in team dynamics, project selection, participant experiences, hackathon settings that align with our research question (RQ1) regarding the influence of the hackathon format on collaboration, competitiveness, and project outcomes. Next, we took a quantitative approach to analyze the types and objectives of the projects of both hackathons to explore how the hackathon formats can influence the types of projects students choose to work on in different hackathon events (RQ2). The study protocol was approved by the author's Institutional Review Board (IRB).

\vspace{-0.06 in}
\subsection{Data Collection for RQ1}
Our data collection process for RQ1 was strategically designed as a combination of semi-structured interviews and on-site observations. Here, we adopted Bandura's Social Learning Theory (SLT) as our theoretical baseline to understand students' perceptions and learning experiences. The principles of SLT emphasize that learning is a social process~\cite{bandura1986social}, shaped by interaction with the environment and others~\cite{maisto1999social, akers2015social, schunk1987peer, vygotsky1978mind, crouch2001peer, smith2009peer}, rather than occurring solely through direct instruction or reinforcement. Since hackathons promote peer learning through observations and collaborations, we considered SLT as our core guideline and used interviews and observational data to enrich our findings.

\subsubsection{Semi-Structured Interviews} 
We conducted semi-structured interviews in RI and GS hackathons. We visited the hackathon venues in person and asked the teams if they would participate in an interview as a team. Teams that agreed to participate were interviewed by two authors of the paper during the event. In these interviews, we asked participants the theme of their project, how they formed their team, how they decided on the project idea, how they distributed the workload of the project, their experience in the event, and their expectations from the event. These interviews provided in-depth insights directly from the participants, allowing us to gain a personal perspective on their experiences. Each interview was audio-recorded and transcribed verbatim for qualitative analysis. On average, each interview took 17.4 minutes to complete and we paid \$15 to each team for their participation.

\subsubsection{On-site observations}
One author (not involved in the interview process) attended both events in person (two days/event) to observe the settings, interactions, and processes of the events. The author talked to the official organizers, event volunteers, and mentors to understand how the groups were located (in terms of locations), how the mentors interacted with the teams, how the workshops were scheduled, and how additional interactive meet-and-greets were planned in each event. In addition, the author moved around and took descriptive documentation of the working culture of project teams without interfering with ongoing activities. The author followed a semi-structured framework for visiting all locations of each event and took notes on anything considered relevant. No audio or video recording was collected. We thematically analyzed the observational notes to examine the interactions, dynamics, and influence of the hackathon settings that may not be fully revealed in the interviews.
\subsection{Data Collection for RQ2}
To answer RQ2, we analyzed the project archives available on the official websites of both GS and RI hackathons (with permission from event organizers). This examination focused on project descriptions, objectives, outcomes, and the roles of participants as documented online. By exploring the details of each project, we categorized them according to their stated objectives, providing a structured way to assess and compare the thematic goals of each hackathon. Additionally, we classified the project teams based on their composition, whether they were regular institutional or gender-specific. This classification was essential for understanding how team gender dynamics might influence the selection of project themes and the overall approach to addressing the hackathon challenges.

\vspace{-0.06 in}
\subsection{Study settings \& Population}

\subsubsection{RI hackathon} The RI 2024 hackathon attracted 343 technology enthusiasts across 101 project teams, with about 26\% mixed-gender teams, 58 all-male teams, and 11 all-female teams. In this study, we interviewed 20 teams consisting of 74 participants (53 males, 19 females, and 2 non-binary), aged 18–25, with diverse gender ratios (11 predominantly male, 2 predominantly female, and 7 mixed-gender), where the mixed gender teams usually had a 3:1 and 2:2 male-to-female ratio. In addition to the core hackathon activities, RI 2024 also featured technical workshops (AI/ML, web, and game development), engaging activities (cup stacking, slideshow karaoke), and networking sessions with industry sponsors, fostering skill development, collaboration, and professional opportunities. The event was open only for students of all skill levels. 

\subsubsection{GS hackathon} 
The GS 2024 attracted 409 participants across 114 teams, specifically targeting underrepresented genders in technology. The event's hybrid format facilitated both on-site and virtual participation. For this study, we interviewed 16 in-person teams (52 participants: 39 female, 13 non-binary, aged 18–25) and four additional female workshop attendees. GS provided diverse workshops (introductory to advanced) for attendees. GS hackathon also featured seven interactive mini-events (e.g., game shows, talent showcases), providing a fun and interactive break from the intense hacking sessions, allowing participants to stay engaged and energized while enjoying a more relaxed and social atmosphere and networking sessions with industry sponsors, promoting skill development, collaboration, and professional connections in an inclusive environment. The event was open to all students who would identify as a marginalized gender.

\vspace{-0.06 in}
\subsection{Data Analysis: Qualitative}
The qualitative analysis began with open coding of interview transcripts and observational notes to identify recurring ideas and participant insights, followed by axial coding to group these into broader themes within a structured analytical framework.
Two coders independently conducted the qualitative coding to ensure the reliability and rigor of our coding process. The coding process began with two coders independently analyzing randomly selected transcripts and observational notes to establish preliminary codes. After collaboratively resolving differences and refining the code list, each coder independently processed the remaining data. Codes were then merged during axial coding to create a cohesive thematic structure. The reliability of this collaborative coding process was quantified using Cohen’s kappa coefficient~\cite{cohen1960coefficient}, which yielded a high agreement level between the coders$(k=0.86, p<.0005)$, underscoring the robustness of our analytic approach. The findings of the qualitative analysis are presented in Section~\ref{quali}.

\vspace{-0.06 in}
\subsection{Data Analysis: Quantitative}
We applied quantitative data analysis to categorize the RI and GS projects. The projects across these two hackathons encompassed a range of categories, including education and professional development, games, finance, accessibility and assistive technology, healthcare, sustainability and lifestyle, entertainment and social experiences, environmental technology, social and creative collaboration, and artificial intelligence systems. Our analysis focused primarily on each hackathon's thematic categorization and lexicon analysis. First, we used BERTopic \cite{grootendorst2022bertopic} to identify topics in each hackathon's project list to get the overarching list of topics. The pre-trained SentenceTransformer model \textit{``all-MiniLM-L6-v2''} \cite{hugging} generated document embeddings, followed by clustering using HDBSCAN. This approach helped us understand the hackathons' most common themes separately and identify different keywords for each topic. Second, we applied a lexicon-based text analysis called Empath \cite{fast2016empath}, works very similarly to LIWC \cite{pennebaker2001linguistic} as they both count category terms in documents, and used it to examine how gender representation shapes thematic focus, innovation priorities, and the societal influences behind project choices. We have discussed our findings from the project analysis in Section \ref{proj}.

%% file: result_sd.tex
\vspace{-0.08 in}
\section{Results}
\vspace{-0.08 in}
\subsection{Outcome of the Qualitative Analysis: Practices and Environmental Factors}
\label{quali}

To answer RQ1, we analyzed interviews and observational notes and found that participants in the GS and RI hackathons follow unique practices. In addition, we observed that the environmental settings of these two hackathons are unique in many ways.

\subsubsection{GS hackathon: Participants' Practices}
At GS hackathon, the composition and formation of teams are carefully designed to address the needs of underrepresented genders in STEM, such as women~\cite{frachtenberg2022underrepresentation, vitores2016trouble}. We found that the teams were gender homogeneous and that many participants (N = 6) described as experiencing an increased level of ``comfort and support'' in that format. A shared sense of identity helped participants establish a deeper camaraderie, where they felt truly supported. They were encouraged to take creative risks and empowered to tackle challenges without hesitation. As one participant shared, \textit{``Being in a team where everyone understands the challenges you face as a woman in tech changes the whole experience for the better''[GS3]}. This team dynamic profoundly reflects participants' perceptions of collaboration, particularly aligning with social learning theory. 

We noticed a larger number of students participating in the GS hackathon were first-time hackathon participants. We sought to understand why participants chose a gender-specific hackathon over traditional regular institutional hackathons for their first participation. Those who had previously attended regular hackathons highlighted that the competitive nature of those events was intimidating, particularly for beginners. Another significant factor was the dominance of male participants, which often led to feelings of marginalization. As GS9 stated, \textit{``In regular institutional hackathons, I often felt overshadowed by male-dominated teams. Finding my footing or feeling like I belonged was hard'' [GS9]}. Participants described GS as a supportive learning environment that allowed them to learn independently. 

Another key factor of the GS hackathon was the mentors. Mentors were assigned to each team, guiding and influencing project selection while ensuring teams retained autonomy. Their support helped participants navigate challenges, stay on track, and take on leadership and technical roles. GS12 stated \textit{``The supportive environment of this GS hackathon pushed my boundaries of what I thought was impossible for myself in this field'' [GS12]}. Inclusive role assignments and strong mentorship ensured that everyone, especially beginners, could actively contribute and grow. This collaborative spirit enhanced inclusivity and learning but sometimes resulted in less defined project outcomes compared to more competitive environments. For instance, some teams struggled to complete their projects on time, often due to varying experience levels and a lower competitive drive. While participants valued the supportive, low-pressure environment, many recognized the need to integrate some competitive elements to encourage more precise, more structured project results while maintaining an empowering and inclusive environment at the GS hackathon. 


\subsubsection{RI hackathon: Participants' Practices}

Unlike the GS, team formation at the RI hackathon was more prevalent based on familiarity and skill, with participants primarily selecting teammates who shared similar interests or technical expertise for project development. While this approach streamlined collaboration, it also contributed to forming predominantly male groups. Our survey of 20 teams revealed that 55\% were male-only, while 35\% were mixed-gender, typically with a 3:1 male-female ratio, and the rest were female-only, reflecting the predominance of male participants. Familiarity influenced team selection, reinforcing traditional gender roles, with women frequently assigned to design or non-technical tasks rather than coding \cite{schauer2025internal}. RI11 highlighted this challenge, stating, \textit{``It's challenging to assert yourself into the more technical roles when men typically assume them, and there's an unspoken assumption about who is better suited for these tasks'' [RI11]}. This dynamic perpetuates gender disparities in technical skills and leadership, echoing broader industry trends. The male-only teams were mostly open to forming a mixed-gendered team if the females had the same interests and the right skill sets. However, we learned a few teams (N = 4) not willing to have female members in their teams, primarily due to two reasons. One, they mentioned they were ``more comfortable'' and ``communication was easier'' with their male friends. And two, females or other marginalized genders ``have alternative options'' since there are hackathons that are solely designed for the minority gender population. Observations and interviews showed that RI hackathon participants selected projects independently, with mentors assisting only upon request. While the competitive environment was seen as rewarding, it often limited collaboration and relationship-building. \textit{``It felt more like a test of skill rather than a place to meet new people''[RI4]}, mentioned [RI4]. While some thrived under pressure, others found it ``stressful'', preferring ``structured networking'' to promote collaboration. Unlike GS, RI’s networking was career-driven rather than based on peer learning. As one participant shared, \textit{``Everyone was focused on making connections that could lead to internships or jobs rather than just learning together''[RI7]}. These findings suggest RI’s competitive atmosphere could benefit from more structured engagement to balance performance with meaningful collaboration.

\subsubsection{GS hackathon: Environmental Factors}
In our analysis, GS's environment mainly focused on learning and collaboration, enhanced by various playful activities that promoted a welcoming, inclusive, and less competitive atmosphere. Participants actively participated in those activities, such as karaoke, dance games, sticker competitions, and other forms of entertainment. This hackathon was hosted in an auditorium that facilitated large gatherings, enhancing its communal aspects. A significant feature of GS hackathon was the relationship between participants and mentors, who were primarily female, reinforcing its gender-centric focus. Participants frequently sought their guidance, easing uncertainties, especially for those new to the hackathon. Additionally, motivational speeches by previous winners played a pivotal role in inspiring participants through success stories and practical advice. This aspect of \textit{"mentorship"} or \textit{"inspirational leadership"} was key in boosting motivation and encouraging participants to participate in more hackathons. Moreover, GS hackathon offered a large number of workshops covering beginner and advanced themes. These workshops were meticulously designed to cater to a broad range of skills and interests, ensuring that all participants could find valuable learning opportunities regardless of their experience level. Overall, the environment at GS hackathon emphasized learning, mentorship, and building community, making it a nurturing space for participants to explore their interests in technology. 

\subsubsection{RI hackathon: Environmental Factors}
At RI, we observed a strong emphasis on skills and achieving tangible results, reflecting a highly competitive, results-driven atmosphere. Multiple workshops covered AI/ML, web development, and game development but catered to those with some prior knowledge, unlike the GS, which offered beginner-friendly sessions. Most RI hackathon participants had prior hackathon experience, unlike GS, reflecting a more skilled demographic. The event’s layout, with teams spread across separate rooms and courtyards throughout the campus, facilitated focused, uninterrupted work but limited cross-team interaction, potentially reducing idea exchange. Similar to GS, RI hackathon included activities like cup stacking and karaoke, offering participants a break from their projects. However, most remained focused on their work, with these activities serving as optional diversions rather than central elements of the event. Unlike the  GS, the mentor engagement level was lower here, possibly due to not having designated mentors for each team and the higher experience levels of the participants not requiring frequent guidance. Overall, the RI hackathon focused on results and advanced skills development, which made it an ideal venue for seasoned developers and students to evaluate their abilities and compete for recognition in a competitive atmosphere.

\vspace{-0.06 in}
\subsection{Outcome of the Quantitative Analysis: Project Topics}
\label{proj}

To examine how the hackathon format influenced the types of projects presented (RQ2), we conducted a two-layered analysis of submissions from the GS and RI hackathons. First, we applied BERTopic for topic modeling, identifying five themes per event. Two authors independently reviewed the topics and keywords, reaching consensus on three common and four unique themes (two per hackathon), with strong inter-rater agreement ($k=0.89$, $p<0.05$). BERTopic also detects outliers; projects that do not fit any cluster. Given the small dataset, we only retained clusters with at least five items and reassigned outliers to the nearest relevant themes, reducing them to under 7\% of all projects. These outliers were excluded from further analysis, as they were not representative. The final themes are summarized in Table~\ref{tab:thematic_analysis}.

To capture linguistic framing, we used Empath to analyze semantic patterns in project names. RI hackathon projects emphasized competition, entertainment, and sustainability, while GS hackathon projects highlighted health, trust, finance, and productivity. These findings show how format shaped both problem framing and solution space.

\begin{figure}
    \centering
    \includegraphics[width=0.9\linewidth]{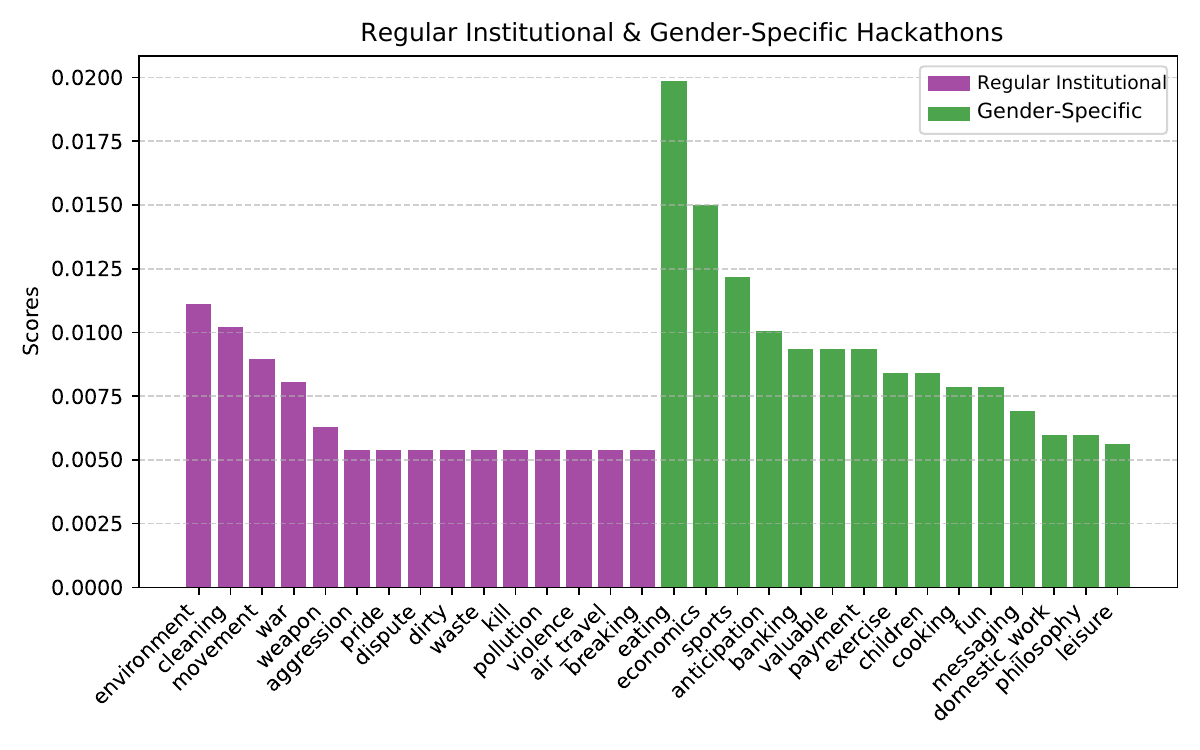}
    \caption{Top 15 unique Empath category scores for RI and GS hackathon}
    \label{fig:ind_hack}
\end{figure}

\vspace{-5pt}
\begin{table}[htbp!]
    \centering
    \footnotesize 
    \renewcommand{\arraystretch}{1.1} 
    \setlength{\tabcolsep}{3pt} 
    \begin{tabular}{p{2.2cm} p{5.8cm}} 
        \toprule
        Event & Category and Keywords \\
        \midrule
        RI hackathon & \textbf{Environmental and Sustainability} (eco, environmental, waste, sorting, sustainability) \\
        & \textbf{Games and Entertainment} (game, run, maze, party, fight, escape, horror) \\
        \midrule
        GS hackathon & \textbf{Productivity and Finance Management} (productivity, expense, budgeting, finance, tracker) \\
        & \textbf{Health and Well-being} (self-care, wellness, balance, fitness, care, emotion) \\
        \midrule
        Both & \textbf{AI-Assistants and Automation} (ai, assistant, voice, virtual, automated, smart) \\
        & \textbf{Education and Learning} (learning, detective, college, internships, eco, domain) \\
        & \textbf{Data Privacy and Awareness} (data, privacy, analysis, training, scams, research) \\
        \bottomrule
    \end{tabular}
    \caption{Categorization of Project Themes Across RI and GS hackathon}
    \label{tab:thematic_analysis}
    \vspace{-0.1 in}
\end{table}
\vspace{-5pt}
BERTopic identified two distinct themes unique to RI hackathon: environmental sustainability and gaming applications. The environmental category encompassed projects addressing waste management and ecological sustainability, reflected in keywords such as "eco," "environmental," and "sorting." Similarly, RI showed a strong emphasis on entertainment and gaming, as evidenced by keywords like "game," "maze," and "fight." Empath further reinforces these findings through lexicon analysis. RI’s environmental focus is quantitatively supported by high scores in "cleaning" (0.003584) and "environment" (0.013799), highlighting the sustainability theme. Meanwhile, the prevalence of terms such as "fight" (0.010215), "competing" (0.005376), and "dispute" (0.005376) confirms RI’s distinct orientation toward interactive and competitive gaming experiences. Projects such as \textit{EcoSort: Smart Waste Management}, \textit{Waste Sorting Adventure}, \textit{Control Space Waste} and \textit{Environmental Game Track} align with these findings, promoting sustainable solutions. Likewise, entertainment-driven projects, including \textit{Maze Runner AI}, \textit{Climate Change Awareness Game}, \textit{Beginner-Friendly Fighting Game} and \textit{Hunger-Games and Mario Party}, exemplify RI’s strong engagement in gamified experiences (Figure \ref{fig:ind_hack}).

In contrast, BERTopic categorized GS’s unique focus on finance, productivity, and well-being. GS hackathon projects frequently aimed to develop practical life management tools, with keywords such as "finance," "expense," and "budgeting", highlighting a strong interest in personal finance solutions. Additionally, themes of health and well-being emerged, as indicated by terms like "self-care," "wellness," and "emotion." Empath supports these thematic findings with dominant lexical categories such as "money" (0.003745), "banking" (0.009363), and "economics" (0.014981), reinforcing GS’s emphasis on financial responsibility. Similarly, the well-being theme is reflected in high scores for "positive\_emotion" (0.005618), "exercise" (0.008427), and "trust" (0.005618). Notable GS projects such as \textit{Expense Tracker Pro}, \textit{Smart Budgeting Assistant}, \textit{Enhance Productivity for Writers}, and \textit{Health Insurance Cost Predictor} align with these trends, prioritizing financial stability. Additionally, health-oriented projects such as \textit{Cooking and Healthy Diet Balance}, \textit{Self-Care Companion}, \textit{Support for Daily Chronic Wellness}, \textit{Self-Care and Balance for College Students} and \textit{Fitness Trackers via Princess} reinforce GS hackathon’s focus on well-being and self-improvement (Figure \ref{fig:ind_hack}).

\begin{figure}
    \centering
    \includegraphics[width=0.8\linewidth]{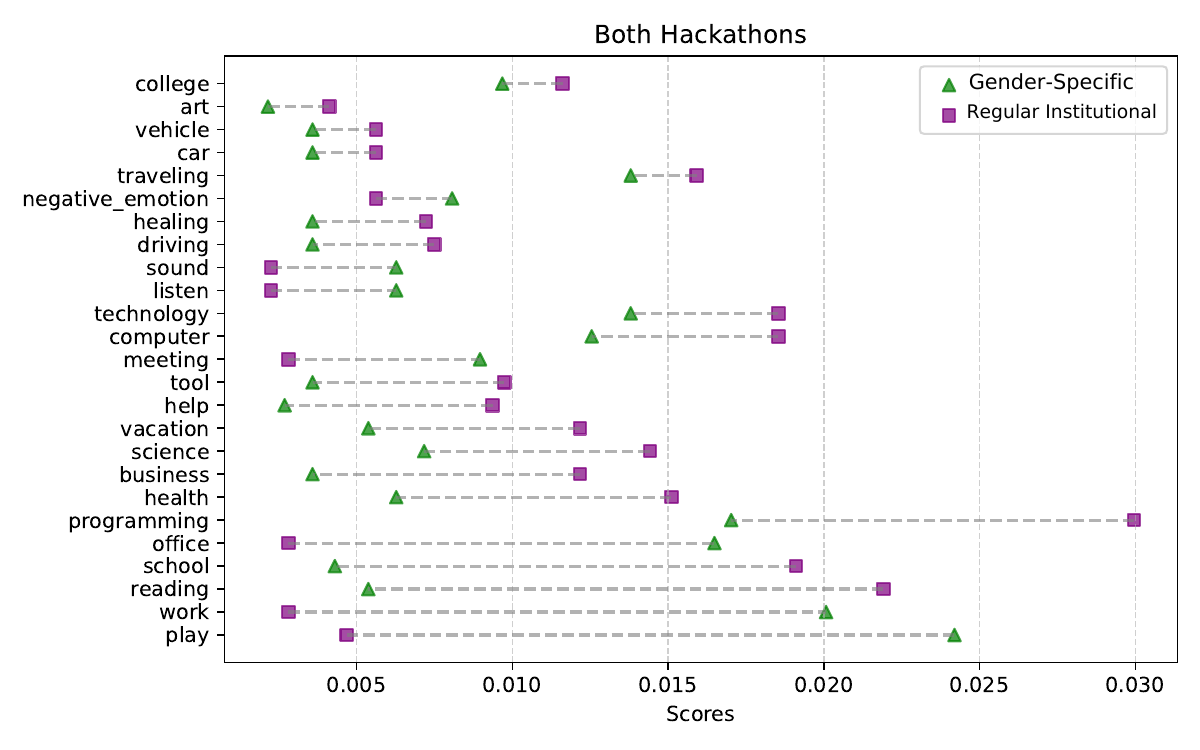}
    \caption{Comparison of top 25 overlapping Empath category scores across RI and GS hackathons}
    \label{fig:both_hack}
    \vspace{-0.08 in}
\end{figure}

Both hackathons exhibit a shared interest in AI assistants, education, and data privacy, as revealed by BERTopic and reinforced by Empath’s lexical analysis. AI-driven solutions appear across various projects such as \textit{AI Stock Prediction (RI)}, \textit{AI-Powered Visual Assistance for the Visually Impaired (GS)}, and \textit{Mood Boards with AI-Powered Assistance (GS)}, highlighting AI’s growing role in practical applications. Epath’s analysis aligns with education and career development, with high lexical scores in "school" (0.004301, 0.019101), "reading" (0.005376, 0.021910), and "college" (0.009677, 0.011610). Projects such as \textit{Study Group Finder (GS)}, \textit{AI-Powered Conversation Skill Builder (GS)}, \textit{Smart Career Fair Assistant (RI)}, \textit{Facilitate Mutual Learning Between Peers (GS)}, and \textit{Internship Finder (RI)} demonstrate AI’s impact on job-seeking, mentorship, networking, and personalized learning. The prevalence of "programming" (0.017025, 0.029963) and "technology" (0.013799, 0.018539) highlights AI’s broader technical influence. Beyond education, data privacy is a key theme, reflecting awareness of digital security, user protection, and ethical AI practices. Projects such as \textit{Privacy Guardian AI (RI)}, \textit{Scam Awareness App (GS)}, \textit{Data Privacy and Security in Research (RI)}, and \textit{Understand the Prevalence of Romance Scams (GS)} address fraud detection and cybersecurity, supported by lexical scores in "privacy" (0.003584, 0.009738), "legal" (0.002688, 0.009363), and "security" (0.005618, 0.012172). The overlapping lexicon values are shown for visualization in Figure \ref{fig:both_hack}.

%% file: Discussion.tex
\vspace{-0.08 in}
\section{Discussion}


The distinct cultures of GS and RI hackathons significantly shaped collaboration and competition. The GS hackathon facilitated inclusivity, teamwork, and leadership, especially among underrepresented groups, while the RI hackathon emphasized technical performance within a competitive, results-driven environment, often at the expense of inclusivity. These differences reflect broader industry dynamics, with  GS hackathon emphasizing empowerment and social impact, and RI hackathon highlighting the competitive nature of the tech sector. A hybrid model that combines GS’s collaborative ethos with RI’s competitive edge could balance motivation, performance, and inclusion. Participant experiences also reveal differing team formation strategies: RI teams prioritized skill alignment and familiarity, whereas GS teams emphasized comfort and affinity. Recognizing these patterns is essential for designing hackathons that promote innovation while supporting diverse and inclusive participation. In addition, our project analysis further highlights gender-influenced preferences in project themes. Previous research shows that when making project choices, individuals favor skills, social bonds, familiarity, and interests, which may explain the distinct selection of projects observed in these hackathons \cite{whalley2017student}. We found fewer health and well-being-focused projects in the RI hackathon compared to GS, where all male groups were more prevalent. Studies indicate that men are generally less likely to discuss or prioritize their health compared to women \cite{obireoziemen, sagar2019improving, sharp2022people}, a topic more openly embraced by female participants at GS hackathon. Also, the female members shared that they often find a lack of effective health-related apps for them. This broader cultural reticence likely contributes to the lower number of health-related projects in regular institutional hackathons. On the other hand, the prevalence of gaming and entertainment projects at RI hackathon, predominantly pursued by all-male teams, suggests a strong male inclination toward gaming and game development \cite{prescott2011segregation, veltri2014gender, forbesGame}. Existing research indicates that women generally exhibit greater environmental concern and engage more in pro-sustainability behaviors than men \cite{li2022female, yaleEnvironment, echavarren2023gender}. However, our study found that the male-dominated RI hackathon featured more environmental projects than GS, which appears to contradict these trends. This discrepancy may be explained by the competitive, impact-driven culture of RI hackathon, which framed environmental sustainability as a high-stakes technical challenge, making it more appealing to participants. On the other hand, the GS’s emphasis on collaboration and inclusivity steered participants toward financial management tools and productivity apps, which align with their practical needs. These projects prioritize accessibility, offer tangible benefits, and are beginner-friendly, enhancing daily life rather than purely technical innovation. In addition, our analysis positions hackathons as intersections of technological innovation and sociocultural influence, where project themes reflect both technical expertise and participant values. Events prioritizing social impact and inclusivity tend to attract diverse participants and foster solutions to broader societal issues. Notably, first-time participants favored the GS hackathon for its welcoming and collaborative atmosphere. To promote accessibility, organizers should enhance outreach workshops that support both newcomers and experienced participants, encouraging diversity in participation and project focus. Here, it is critical to understand that although both hackathons were organized by public institutions, the institutions varied in scale and resource availability, which may have some impact on environmental factors and practices reported by this project. Future work needs to carefully consider this aspect to accurately interpret our findings. 


\vspace{-0.1 cm}
\section{Limitations}
 

While our findings offer valuable insights, several limitations need to be considered when interpreting the results. The modest sample size, particularly for qualitative interviews and project analysis, may not fully capture the diversity of global hackathon experiences, highlighting the need for broader participant groups in future research. Additionally, both hackathons were geographically concentrated within the United States in public institutions, suggesting the importance of exploring varied organizational and cultural contexts to better reflect global hackathon dynamics. These limitations contextualize our findings and underscore opportunities for further investigation.

\vspace{-0.1 in}
\section{Conclusion and Future works}
This study examined the dynamics of gender-specific and regular institutional hackathons, revealing distinct thematic and environmental differences. Gender-specific events often emphasized community-centered themes such as health, well-being, and personal finance, while regular hackathons leaned toward entertainment and environmental sustainability, reflecting a more competitive, technically driven focus. These findings suggest that hackathons function not only as sites of innovation but also as reflections of participants’ values and interests. Our results point to the dual role of hackathons in fostering both collaboration and competition, and we propose that hybrid models may better support inclusive participation while retaining competitive motivation. Future work could investigate how theme selection shapes participants’ long-term engagement with computing and their professional trajectories.



\newpage